\documentclass[a4paper,11pt]{article}
\usepackage{amsmath}
\usepackage{amssymb}
\usepackage{graphicx}
\usepackage{cite}
\usepackage[T1]{fontenc}
\usepackage[a4paper,left=1.0in, right=1.0in,top=1.1in, bottom=1.2in]{geometry}
\newcommand{\be}{\begin{equation}}
\newcommand{\ee}{\end{equation}}
\begin{document}
\titlepage
\vspace*{1in}
\begin{center}
{\Large \bf Hard scale dependent gluon density, saturation and forward-forward
dijet production at the LHC.}\\
\vspace*{0.5cm}
Krzysztof Kutak \\
\vspace*{0.5cm}
 {\it Instytut Fizyki Jadrowej im H. Niewodniczanskiego,\\
Radzikowskiego 152, 31-342 Krakow, Poland\\}
\end{center}
\vspace*{1cm}
\centerline{(\today)}

\vskip1cm
\begin{abstract}
We propose a method to introduce Sudakov effects to the unintegrated gluon density promoting it to be hard scale dependent. The advantage of the approach is that it guarantees that the gluon density is positive definite and that the Sudakov effects cancel on integrated level. As a case study we apply the method to calculate angular correlations and $R_{pA}$ ratio for p+p vs. p+Pb collision in the production of forward-forward dijets.  
\end{abstract}
\section*{Introduction}
In perturbative QCD the theoretical construction that is used to evaluate cross sections for hadron hadron collisions
is a factorization \cite{Collins:2011zed} which allows to split the cross section into parton densities characterizing the incoming hadrons and hard subprocess. In particular high energy factorization \cite{Gribov:1984tu,Catani:1990eg} is a prescription for such a decomposition which allows for taking into account off-shellnes of incoming partons carrying low longitudinal momentum fraction of hadrons $x$ already at the lowest order accuracy both in matrix elements \cite{vanHameren:2012if,vanHameren:2012uj,Kotko:2014aba,vanHameren:2014iua,Nefedov:2013ywa,Maciula:2013kd,Lipatov:2014mja,Lipatov:2014yna} and parton densities. Its applications to situations where saturation effects are relevant is a phenomenologically useful. There are already results which generalize it in some limit of phase space to include saturation \cite{Dominguez:2011wm}.
The basic ingredient of the formula for factorization is the unintegrated gluon density. In the high energy limit it comes from resummation of emission of gluons emitted in the $s$ channel which are ordered in the longitudinal momentum fractions and unordered in the transversal momenta.      
When the longitudinal momentum fraction is $x\!<\!<\!1$ one argues that the nonlinear effects start to show up to tame the rapid power like growth of the gluon density \cite{Gribov:1984tu} and there are indications that indeed saturation occurs in nature \cite{GolecBiernat:1998js,Stasto:2000er,Albacete:2010pg,Dumitru:2010iy}. Resummation of relevant contributions for introducing unitarity corrections can be achieved conveniently in the coordinate space in the so called dipole picture (virtual probe interacting with target is represented as a color dipole) and leads to the Balitsky-Kovchegov (BK) equation \cite{Balitsky:1995ub,Kovchegov:1999yj,Kovchegov:1999ua} or its generalizations \cite{Kovner:2014lca}. 
The unintegrated gluon density  (also called dipole gluon density) used in the high energy factorization  can be obtained as the Fourier transform of a dipole amplitude $N$
\be
{\cal F}(x,k^2)=\frac{C_F}{\alpha_s(2\pi)^3}\int  d^2 {\bf b} d^2{\bf r}e^{-i{\bf k}\cdot{\bf r} }\nabla^2_r N({\bf r},{\bf b},x)
\ee
where $N$ is the solution of the coordinate space BK equation and ${\bf b}$ is the impact parameter at which a color dipole collides with the target,
the size of a dipole is $r=|{\bf r}|$ and $k=|{\bf k}|$ is momentum. The two dimensional vectors  
$|{\bf r}|$ and $|{\bf k}|$ lay in the transversal plane to the collision axis.

It turns out that in order for the BK equation to be applicable for processes at LHC one necessarily needs to include resummed corrections of higher orders among which the kinematical constraint \cite{Kwiecinski:1996td,Andersson:1995jt} is the most dominant. It softens the singularities of the evolution kernel and therefore slows down the evolution. Its inclusion in the BFKL kernel allows for a reasonable well description of $p_T$ spectra of forward-central dijet at the LHC \cite{Kutak:2012rf,Kotko1}.
Another type of effect that is beyond the BK is the angular ordering leading to dependence
of the gluon density on the scale of the hard process. At the linear level, inclusion of  such effects leads to the CCFM evolution equation\footnote{recently fitted to $F_2$ data in \cite{Hautmann:2013tba}}\cite{Ciafaloni:1987ur,Catani:1989sg,Catani:1989yc} while at the nonlinear level to equations introduced in \cite{Kutak:2011fu,Kutak:2012yr,Kutak:2012qk}. Importance of the hard scale dependence has been also recognized  by \cite{Kimber:2001sc,Kimber:1999xc} where the effects of coherence were introduced in the last step of evolution. The later framework is particularly interesting as it is relatively straightforward to apply since it uses parton densities which might come from a collinear framework   on top of which the Sudakov  effects are applied \cite{Sudakov:1954sw} in a factorized form are applied.  
Furthermore, it has been noticed in \cite{vanHameren:2014ala} that in order to obtain description of data in a wider domain of the $\Delta\phi$ one needs to include Sudakov effects in the low-$x$ framework to  ensure no emissions between the scale $k$ of the gluon transverse momentum, and the scale $\mu$ of the hard process. In the method described in \cite{vanHameren:2014ala} the Sudakov effects were imposed on the cross section level i.e. generated events were weighted with a Sudakov form factor preserving unitarity, assuring that the total cross section will not be affected \footnote{The method used in \cite{vanHameren:2014ala} will be soon available within LxJet program  	\cite{kotko}}. 
Another approach to introduce Sudakov effect is to include them directly as a part of the evolution equation i.e. at all steps in the evolution. Such an approach leads to the already mentioned CCFM evolution and the Sudakov form factor gets the interpretation of an object which resums virtual and unresolved real corrections relevant when the scale of the harder process is larger than the local 
$k$ of the gluon.\\
In the present paper we start directly from the gluon density summing low-$x$ logarithms,  accounting for nonlinearities and promote it to depend on the hard scale. This method is attractive since it provides gluon density which, once constructed, can be used in various phenomenological applications.
We perform our construction for proton and lead and apply the resulting gluon density to provide estimates of relevance of coherence for the nuclear modification ratio $R_{pA}$ in the production of forward-forward dijet. 
\section*{Sudakov effects and unintegrated gluon density}
The solutions of the evolution equation combining physics of saturation and coherence  show that saturation scale gets nontrivial dependence on the scale of the hard process \cite{Deak:2012mx,Kutak:2013yga} and leads for instance to the effect called saturation of the saturation scale \cite{ Avsar:2010ia}. Due to its numerical complexity the equation has not been still applied to phenomenology. Below we propose a model prescription how to introduce hard scale dependence on top of the unintegrated gluon density ${\cal F} (x,k^2)$ 
obtained from solutions of BK or BFKL evolution equation.\\
The prescription is motivated by the method developed in \cite{vanHameren:2014ala} but formulated in terms of the unintegrated gluon density. Therefore the methods are formally not equivalent. The comparison of the methods is postponed for future studies.
The basic assumptions are the following: 
\begin{itemize}
\item
on the integrated level the gluon densities obtained from the hard scale dependent gluon density ${\cal F}(x,k^2,\mu^2)$ and ${\cal F}(x,k^2)$ are the same. This guarantees that the Sudakov form factor just modifies the shape of the gluon density but on the inclusive level the distribution is the same.
\item Contribution with $k>\mu$ is given by the unintegrated gluon density ${\cal F}(x,k^2)$ which could be obtained by solving the BK equation. 
\end{itemize}
The assumptions above lead to the following formula:
\be
{\cal F}(x,k^2,\mu^2):=\theta(\mu^2-k^2)T_s(\mu^2,k^2)\frac{xg(x,\mu^2)}{xg_{hs}(x,\mu^2)}{\cal F}(x,k^2)+\theta(k^2-\mu^2){\cal F}(x,k^2).
\label{eq:hardgluon}
\ee

where 
\be
xg_{hs}(x,\mu^2)=\int^{\mu^2} dk^2 T_s(\mu^2,k^2){\cal F}(x,k^2),\,\, xg(x,\mu^2)=\int^{\mu^2} dk^2{\cal F}(x,k^2)
\label{eq:intglu2}
\ee
and the Sudakov form factor assumes the form:
\be
T_s(\mu^2,k^2)=\exp\left(-\int_{k^2}^{\mu^2}\frac{dk^{\prime 2}}{k^{\prime 2}}\frac{\alpha_s(k^{\prime 2})}
{2\pi}\sum_{a^\prime}\int_0^{1-\Delta}dz^{\prime}P_{a^\prime a}(z^\prime)\right)
\ee
where $\Delta=\frac{\mu}{\mu+k}$ and $P_{a^\prime a}$ is a splitting function with subscripts $a^\prime a$ specifying the type of transition. In the $gg$ channel one multiplies  $P_{gg}(z)$ by $z$ due to symmetry arguments \cite{Kimber:2001sc}.

\begin{figure}[t!]
\begin{picture}(30,30)
   \put(30, -110){
      \includegraphics{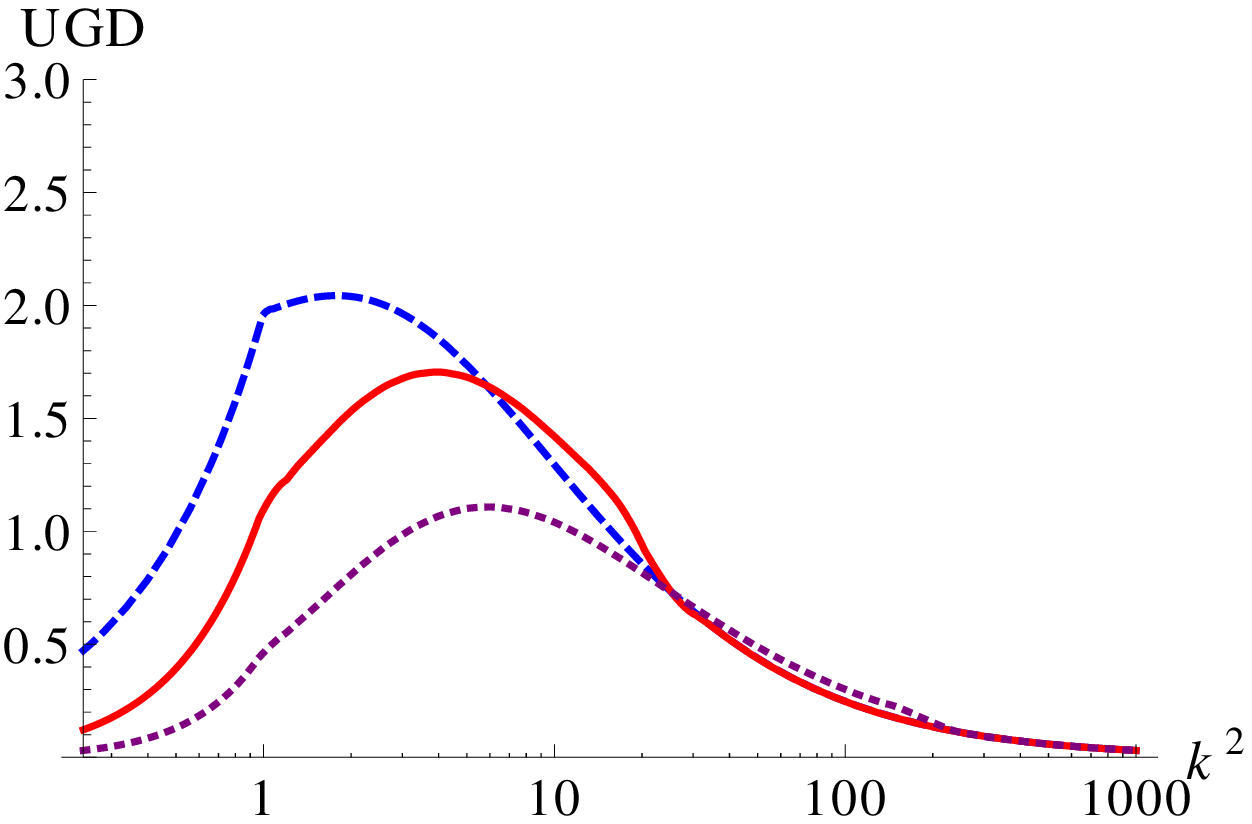}
    }

\put(280, -110){
      \includegraphics{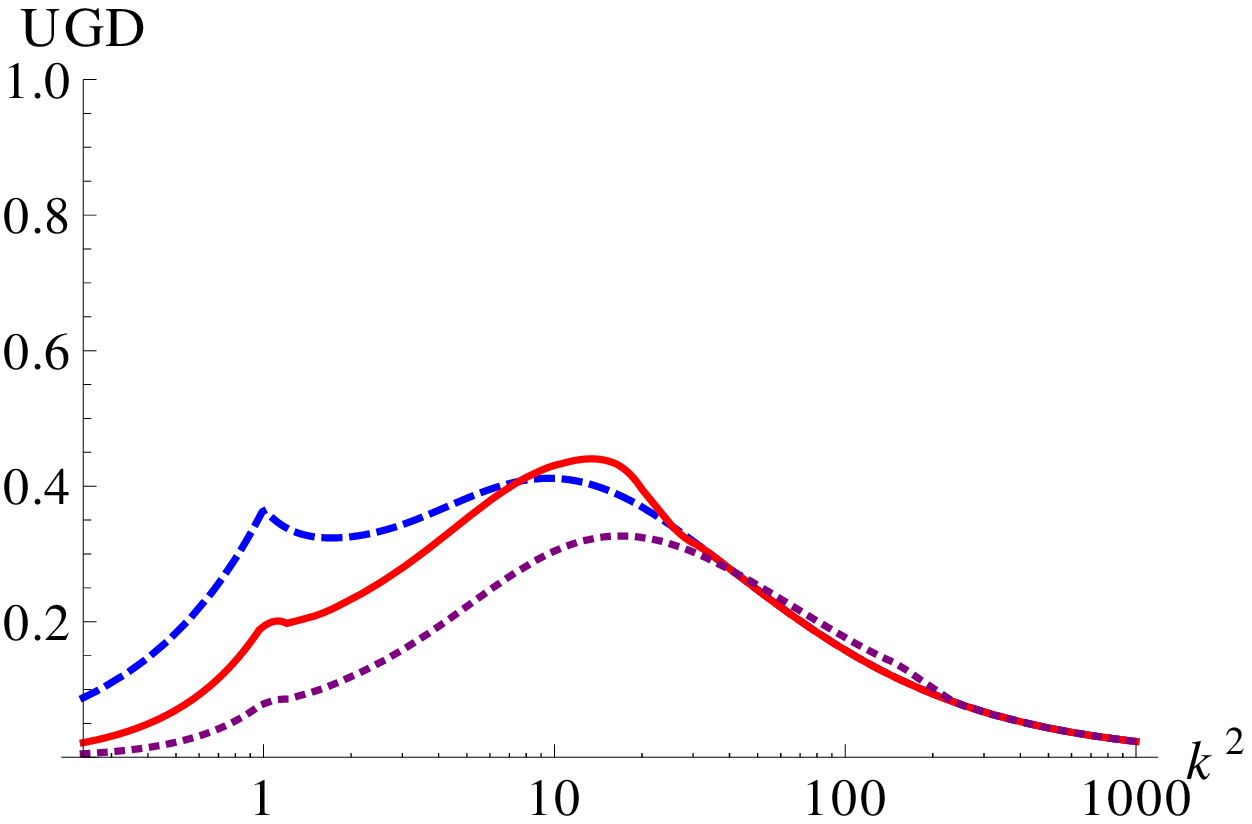}
    }

\end{picture}
\vspace{4cm}
\caption{\small We abbreviate unintegrated gluon density by UGD. 
Left: unintegrated gluon density of proton with Sudakov effects evaluated at $x=10^{-5}$ at hard scale $\mu^2=20GeV^2$ (continuous red line), hard scale $\mu^2=200GeV^2$ (purple dotted line), unintegrated gluon density without Sudakov effects evaluated at $x=10^{-5}$ (blue dashed line). Right: unintegrated gluon density of Pb with Sudakov effects evaluated at $x=10^{-5}$ at hard scale $\mu^2=20GeV^2$ (continuous red line), hard scale $\mu^2=200GeV^2$ (purple dotted line), unintegrated gluon density without Sudakov effects evaluated at $x=10^{-5}$ (blue dotted line) }
\vspace{0.5cm}
\label{fig:plot1}
\end{figure}

The construction guarantees that at the integrated level the number of gluons does not change, 
since after integration up to the hard scale in Eq. (\ref{eq:hardgluon}) and application of Eq. (\ref{eq:intglu2}) the terms $xg_{hs}$ cancel and the part with $\theta(k^2-\mu^2)$ drops.
The Sudakov form factor just makes the shape of the gluon density scale dependent but does not modify its integral.\\
In order to study properties of the introduced hard scale dependent unintegrated gluon density we use the gluon density obtained from the momentum space version of the BK equation in the large target approximation. At leading-order in $\alpha_s \ln(1/x)$ it reads: 
\begin{multline} 
{\cal F}(x,k^2) \; = \; {\cal F}^{(0)}(x,k^2) \\
+\,\frac{\alpha_s(k^2)N_c}{\pi}\int_x^1 \frac{dz}{z} \int_{k_0^2}^{\infty}
\frac{dl^2}{l^2} \,   \bigg\{ \, \frac{l^2{\cal F}(\frac{x}{z},l^2)\,   -\,
k^2{\cal F}(\frac{x}{z},k^2)}{|l^2-k^2|}   +\,
\frac{k^2{\cal F}(\frac{x}{z},k^2)}{|4l^4+k^4|^{\frac{1}{2}}} \,
\bigg\} \\
-\frac{2\alpha_s^2(k^2)}{R^2}\left[\left(\int_{k^2}^{\infty}\frac{dl^2}{l^2}{\cal F}(x,l^2)\right)^2
+{\cal F}(x,k^2)\int_{k^2}^{\infty}\frac{dl^2}{l^2}\ln\left(\frac{l^2}{k^2}\right){\cal F}(x,l^2)\right]\,,
\label{eq:fkov} 
\end{multline} 
where $R$ is the radius of the hadronic target and ${\cal F}^{(0)}(x,k^2)$ is starting distribution.\\
The linear part of (\ref{eq:fkov}) is given by the BFKL kernel while the nonlinear part is proportional to the triple pomeron vertex \cite{Bartels:1994jj,Bartels:2007dm} which allows for the recombination of gluons.
In reference \cite{vanHameren:2014lna} it has been shown that in order to apply the BK equation to dijet physics one necessarily has to go beyond the equation with just running coupling corrections included i.e. the rcBK \cite{Albacete:2010sy}. Therefore to be realistic with applications for LHC we use the momentum space BK equation with corrections formulated in \cite{Kwiecinski:1997ee,Kutak:2003bd,Kutak:2004ym}. 
Those corrections include 
\begin{itemize}
\item kinematic effects limiting the $l$ integration  enforcing the virtuality of exchanged $t$ channel gluon to be dominated by its transversal component. 
\item running coupling
\item pieces of splitting function subleading at low-$x$ important at larger values of splitting ratio $z$ and contribution of sea quarks (indicated below by $\Sigma(x, k^2)$)
\end{itemize}
The final equation assumes the form:
\begin{multline} 
{\cal F}(x,k^2) \; = \; {\cal F}^{(0)}(x,k^2) \\
+\,\frac{\alpha_s(k^2) N_c}{\pi}\int_x^1 \frac{dz}{z} \int_{k_0^2}^{\infty}
\frac{dl^2}{l^2} \,   \bigg\{ \, \frac{l^2{\cal F}(\frac{x}{z},l^2) \, \theta(\frac{k^2}{z}-l^2)\,   -\,
k^2{\cal F}(\frac{x}{z},k^2)}{|l^2-k^2|}   +\,
\frac{k^2{\cal F}(\frac{x}{z},k^2)}{|4l^4+k^4|^{\frac{1}{2}}} \,
\bigg\} \\  + \, \frac{\alpha_s (k^2)}{2\pi k^2} \int_x^1 dz \,
\Bigg[\left(P_{gg}(z)-\frac{2N_c}{z}\right) \int^{k^2}_{k_0^2} d l^{
2}\,{\cal F}\left(\frac{x}{z},l^2\right)+zP_{gq}(z)\Sigma\left(\frac{x}{z},k^2\right)\Bigg] \\
-\frac{2\alpha_s^2(k^2)}{R^2}\left[\left(\int_{k^2}^{\infty}\frac{dl^2}{l^2}{\cal F}(x,l^2)\right)^2
+
{\cal F}(x,k^2)\int_{k^2}^{\infty}\frac{dl^2}{l^2}\ln\left(\frac{l^2}{k^2}\right){\cal F}(x,l^2)\right]\,,
\label{eq:fkovres} 
\end{multline} 
where the input gluon density is given by
\be
{\cal F}^{(0)}(x,k^2)= \frac{\alpha_s(k^2)}{k^2}\int_x^1P_{gg}(z)\frac{x}{z}g(\frac{x}{z},k_0^2),\;\;\;\; xg(x,k_0^2=1)=0.994(1+82.1\,x)^{18.6}.
\ee
The plots of the gluon density obtained from solving (\ref{eq:fkovres}) and its extension for Pb target is shown on Fig. (\ref{fig:plot1}).
The blue lines correspond to the situation when the hard scale effects are not taken into account. The kink at $k_0=1GeV^2$ is an artifact of the matching condition between the model extension (needed for numerical purposes) below $k\!<\!k_0=1GeV$ and the evolution for $k\!>\!k_0=1GeV$.
In the region below  $k\!<\!k_0=1GeV$ which can not be accessed with the used numerical framework the gluon density is assumed to behave like ${\cal F}\sim k^2$.
The maximum of the distribution signals the emergence of the saturation scale.   
The gluon density from equation (\ref{eq:fkovres}) has been successfully applied to the description of $F_2$ structure function data \cite{Kutak:2012rf} and after accounting for Sudakov effects (at cross section level) for the description of azimuthal angle correlations of forward-central dijets in inclusive and inside jet tag scenario \cite{vanHameren:2014ala}. 

\section*{Applications}
In this section we apply the hard scale dependent gluon density to study angular correlations of forward-forward dijet and to calculate the $R_{pA}$ ratio for p+Pb collision \footnote{The calculation has been done within Mathematica package MATH4JET available form the author on request.}.
As argued in \cite{vanHameren:2014lna} this observable is particularly interesting for testing low $x$ effects since the kinematical configuration of two  jets probes the gluon density at $x\approx 10^{-5}$. Furthermore, the distance in rapidity of produced jets is small, therefore the phase space for emission of further jets is suppressed. 
In order to calculate the cross section we are after we use the hybrid high energy factorization\cite{Deak:2009xt}:
\be
  \frac{d\sigma}{dy_1dy_2dp_{1t}dp_{2t}d\Delta\phi} 
  =
  \sum_{a,c,d} 
  \frac{p_{t1}p_{t2}}{8\pi^2 (x_1x_2 S)^2}
  |\overline{{\cal M}_{ag\to cd}|}^2
  x_1 f_{a/A}(x_1,\mu^2)\,
  {\cal F}_{g/B}(x_2,k^2,\mu^2)\frac{1}{1+\delta_{cd}}\,,
  \label{eq:cs-main}
\ee
with  $k^2 = p_{t1}^2 + p_{t2}^2 + 2p_{t1}p_{t2} \cos\Delta\phi$ and
$$
  x_1 = \frac{1}{\sqrt{S}} \left(p_{t1} e^{y_1} + p_{t2} e^{y_2}\right)\,,
  \qquad
  x_2 = \frac{1}{\sqrt{S}} \left(p_{t1} e^{-y_1} + p_{t2} e^{-y_2}\right)\,,
  \label{eq:x1x2}
$$

\begin{figure}[t!]
\begin{picture}(30,30)
   \put(30, -110){
      \includegraphics{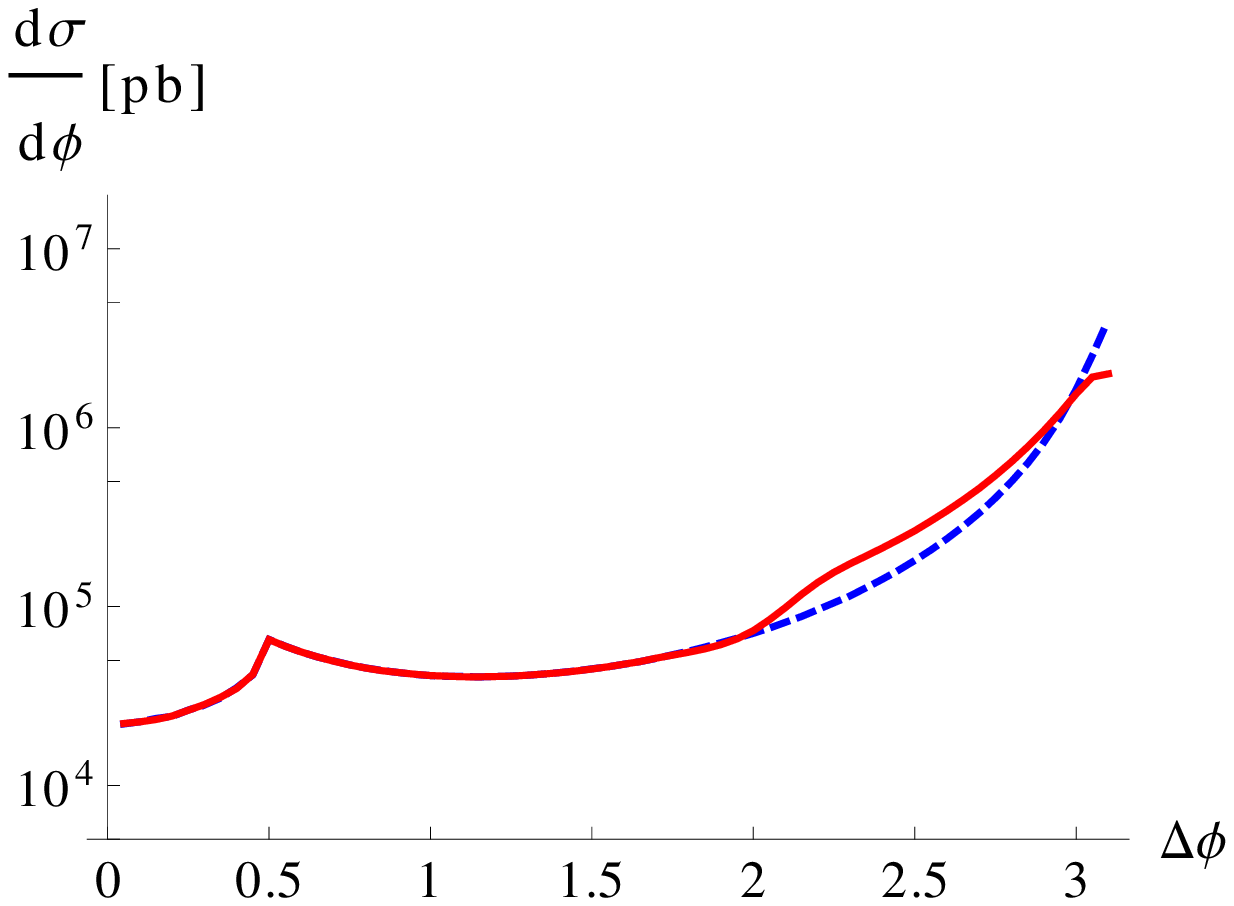}
    }

\put(280, -110){
      \includegraphics{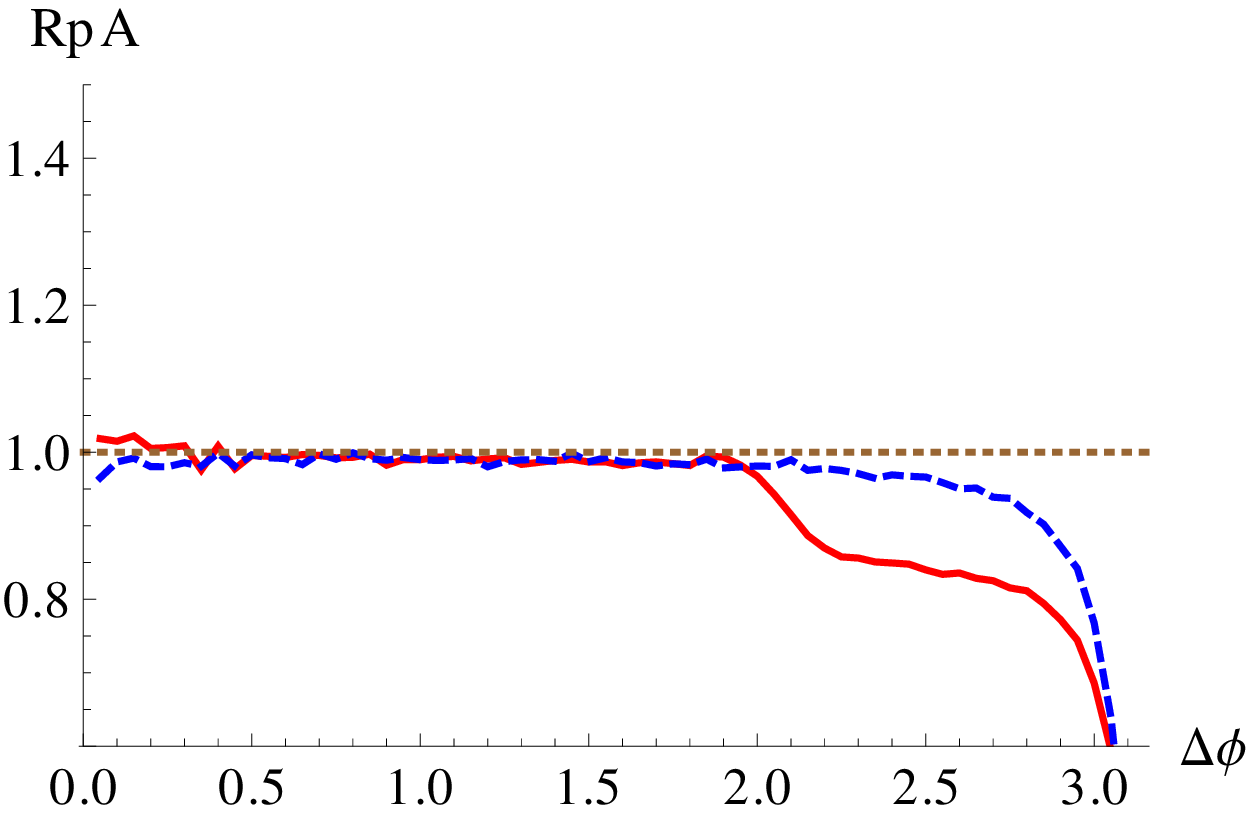}
    }

\end{picture}
\vspace{4cm}
\caption{\small Left: cross section for decorelations in production of forward-forward dijet in p+p collision at $7\,TeV$. The rapidities of produced jets satisfy $p_{t1}\!>\!p_{t2}>20 GeV$. The continuous red line corresponds to the situation with Sudakov effects included while the blue dashed line omits Sudakov effects. Right: the $R_{pA}$ ratio for p+p v. p+Pb. The continuous red line corresponds to situation with Sudakov effects included while the blue dashed line omits Sudakov effects, the brown line just helps to see the deviation from unity.}
\vspace{0.5cm}
\label{fig:plot2}
\end{figure}

In the formulas above $S$ is the squared energy in the center of mass system of the incoming
hadrons (for p+p energy is 7 TeV while for p+Pb it is 5.02TeV) the matrix elements correspond to processes $qg^*\to qg$, $gg^*\to gg$, $gg^*\to \bar{q} q$ and $f(x_1,\mu^2)$ is a collinear parton density while the hard scale is given by $\mu=(p_{t1}+p_{t2})/2$.\\
To visualize the role of the Sudakov effect we calculate the cross section for angular correlations of produced jets Fig.(\ref{fig:plot2}). The kinematical cuts are $p_{t1},\,p_{t2}\!>\!20 GeV$, $4.9\!>\!y_1,\,y_2\!>\!3.2$. and we use the jet algorithm in a form $R=\sqrt{(\Delta y)^2+(\Delta\phi)^2}\!>\!0.5$ i.e. if the distance between two partons is larger than $R$ the partons form two jets while if the distance between $R$ is smaller than $R$ the events are rejected.
The jet algorithm serves as regulator of the collinear singularity of off-shell matrix element which arises when at small rapidity distance the azimuthal angle between produced partons is small.
We see that the Sudakov effect suppresses the cross section when the jets are close to back-to-back configurations while it enhances the cross section in the a region dominated by configurations where the hard scale of the process is a bit larger than the $k$ of the incoming off-shell gluon (similar effect has been observed earlier in studies of forward-central jets in \cite{vanHameren:2014ala}). What is novel is that now one can attribute the enhancement phenomenon to the hard scale dependent gluon density dominating over the regular gluon density at regions where the hard scale is approaching $k$ as seen on Fig.(\ref{fig:plot1}).
The kink visible at small values of $\Delta\phi$ is due to, the jet definition which introduces sharp cut-off of the events not classified as jets. 

\begin{figure}[t!]
\begin{picture}(30,30)
   \put(10, -110){
      \includegraphics{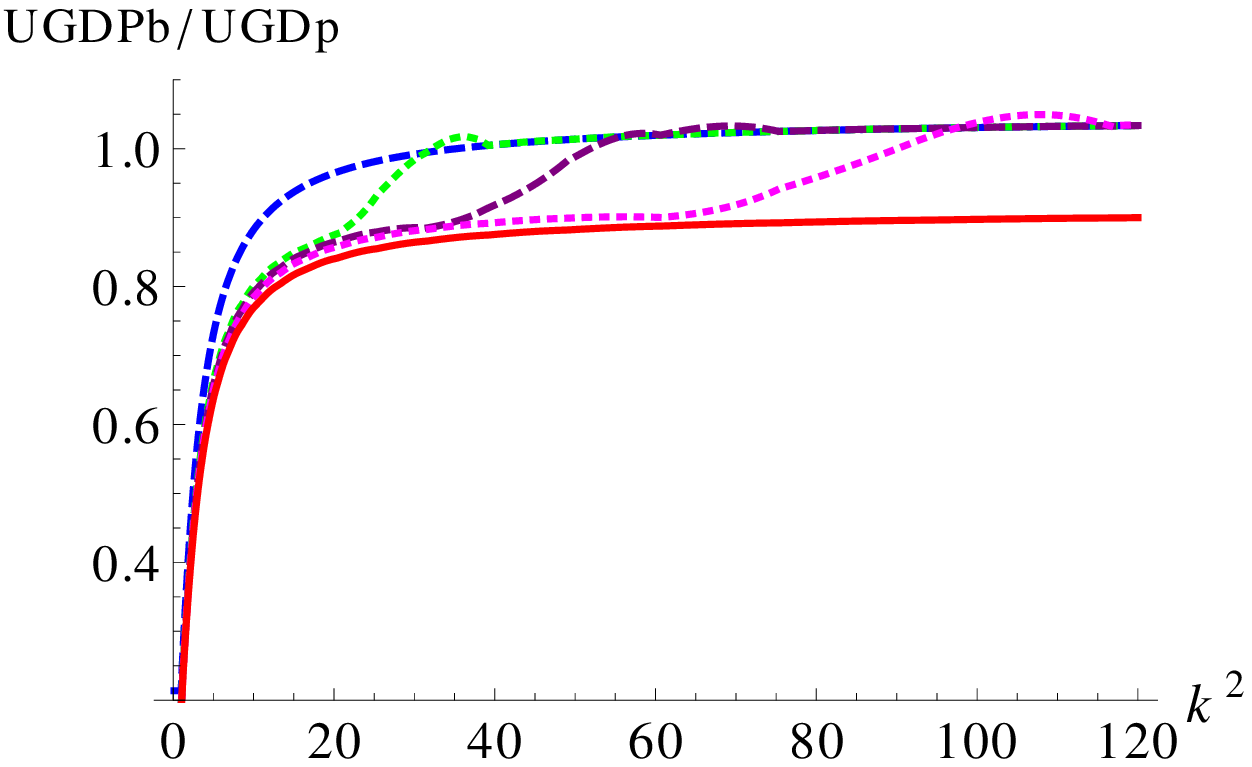}
    }

\put(250, -110){
      \includegraphics{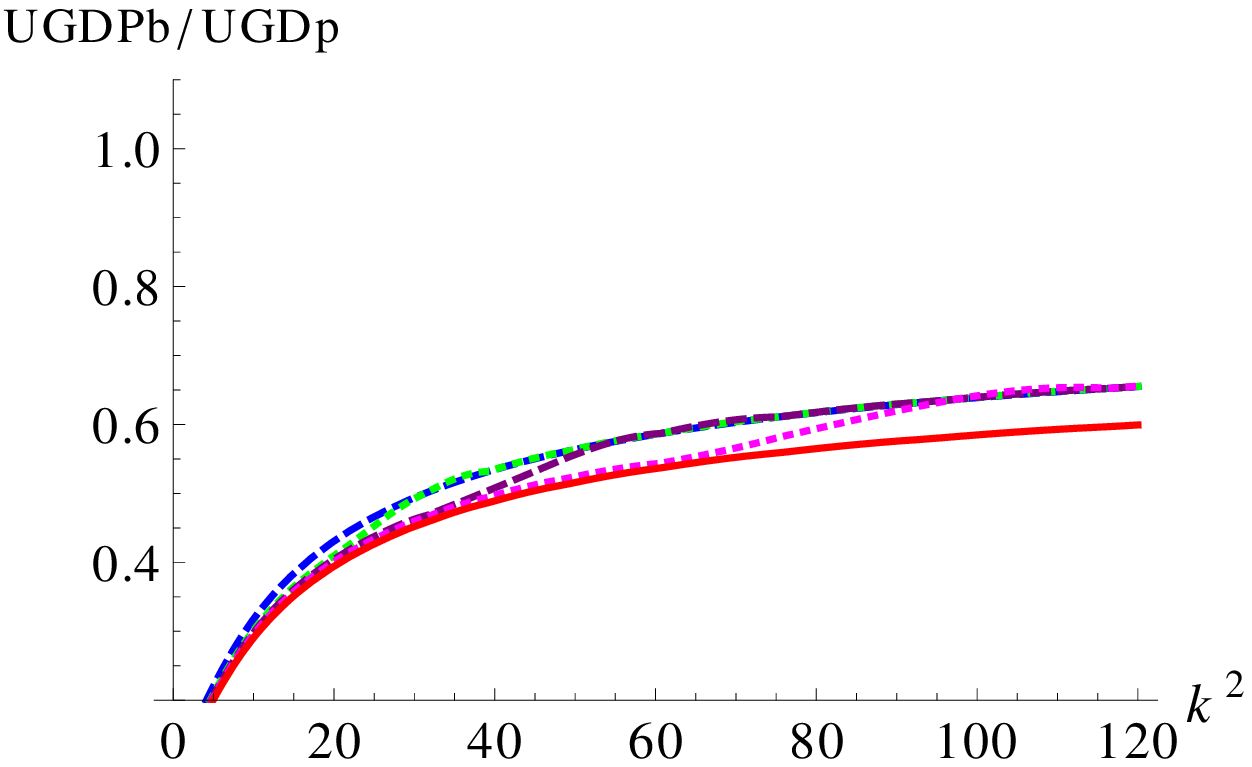}
    }

\end{picture}
\vspace{4cm}
\caption{\small Left: ratio of unintegrated gluon density (UGD) of lead to unintegrated gluon density of proton evaluated at $x=10^{-3}$ at hard scale $\mu^2=25GeV^2$(green dotted line), $\mu^2=45GeV^2$ (purple dashed line),  $\mu^2=80GeV^2$ (magenta dotted line), $\mu^2=400GeV^2$ (red continuous line), no hard scale dependence (blue dashed line). Right: ratio of gluon density of lead to gluon density of proton evaluated at $x=10^{-5}$ at hard scale $\mu^2=25GeV^2$(green dotted line), $\mu^2=45GeV^2$ (purple dashed line),  $\mu^2=80GeV^2$ (magenta dotted line), $\mu^2=400GeV^2$ (red continuous line), no hard scale dependence (blue dashed line)}
\vspace{0.5cm}
\label{fig:plot9}
\end{figure}

To finalize our study we investigate the $R_{pA}$ i.e. the ratio of the cross section for decorelations of dijet produced in p+p and p+Pb. We see that the hard scale dependence leads to the ratio of considered cross sections to be smaller where the saturation effects play a role i.e. at values of large $\Delta\phi$. By inspecting the plots of unintegrated gluon densities and their ratios in Fig. (\ref{fig:plot9}) we see that the gluon density of proton is more affected by Sudakov effects than the lead gluon density. Therefore the ratio is smaller than one in a wider range of $k$ and therefore in a larger range of $\Delta\phi$. This is because in case of lead, the saturation effects are larger and the suppression of low $k$ region is more significant already for, hard scale independent gluon density. \\

\section*{Acknowledgments}
First of all I would like to thank Piotr Kotko for stimulating discussions and comments on the model introduced in this paper. I would like to thank Andreas van Hameren and Edith Smith for proofreaing.\\
Furthermore the stimulating discussions with Andreas van Hameren, Cyrille Marquet, Sebastian Sapeta, Antoni Szczurek are kindly acknowledged. I also would like to thank Rafal Maciula and Mikhail Ryskin for useful correspondence.\\
This work is partly supported by the Polish National Science Center Grant No.  DEC-2011/03/B/ST2/02632 and partly by grant NCBiR Grant No. LIDER/02/35/L-2/10/NCBiR/2011

\end{document}